\documentclass[12pt]{article}
\usepackage[english]{babel}
\usepackage[utf8x]{inputenc}
\usepackage{amsmath}
\usepackage{graphicx}
\usepackage{etoolbox}
\usepackage{changepage}
\usepackage{titlesec}
\usepackage[margin=1in]{geometry}
\usepackage{times}
\usepackage{float}
\usepackage[numbers,super]{natbib}
\usepackage{ragged2e} 

\usepackage{amsmath,amssymb,amsfonts}
\usepackage{algorithmic}
\usepackage{graphicx}
\usepackage{textcomp}
\usepackage{xcolor}
\usepackage{wrapfig}
\usepackage{textcomp}
\usepackage{xcolor}
\usepackage{soul}
\usepackage{braket}

\justifying

\title{\large \textbf{Quantum Computing and Cybersecurity Education: A Novel Curriculum for Enhancing Graduate STEM Learning}} 
\author{\normalsize Suryansh Upadhyay\\
\normalsize sju5079@psu.edu\\
\normalsize Electrical Engineering\\\
\normalsize The Pennsylvania State University
\date{} 
\and
\normalsize Koustubh Phalak\\
\normalsize krp5448@psu.edu\\
\normalsize Computer Science \\\
\normalsize The Pennsylvania State University
\and
\normalsize Jungeun Lee\\
\normalsize jpl5372@psu.edu\\
\normalsize Social Science Research Institute\\\
\normalsize The Pennsylvania State University
\and
\normalsize Kathleen Mitchell Hill\\
\normalsize kmm173@psu.edu\\
\normalsize Centre for Science and the Schools\\\
\normalsize The Pennsylvania State University
\and
\normalsize Swaroop Ghosh\\
\normalsize szg212@psu.edu\\
\normalsize Electrical Engineering and Computer Science\\\
\normalsize The Pennsylvania State University}

\makeatletter 
\patchcmd{\@maketitle}{\begin{center}}{\begin{adjustwidth}{0.5in}{0.5in}\begin{center}}{}{}
\patchcmd{\@maketitle}{\end{center}}{\end{center}\end{adjustwidth}}{}{}
\makeatother

\begin{document}

\justifying
\maketitle

\section*{Abstract}
Quantum computing is an emerging paradigm with the potential to transform numerous application areas by addressing problems considered intractable in the classical domain. However, its integration into cyberspace introduces significant security and privacy challenges. The exponential rise in cyber-attacks, further complicated by quantum capabilities, poses serious risks to financial systems and national security. The scope of quantum threats extends beyond traditional software, operating system, and network vulnerabilities, necessitating a shift in cybersecurity education. Traditional cybersecurity education, often reliant on didactic methods, lacks hands-on, student-centered learning experiences necessary to prepare students for these evolving challenges. There is an urgent need for curricula that address both classical and quantum security threats through experiential learning. In this work, we present the design and evaluation of EE-597: Introduction to Hardware Security, a graduate-level course integrating hands-on quantum security learning with classical security concepts through simulations and cloud-based quantum hardware. Unlike conventional courses focused on quantum threats to cryptographic systems, EE-597 explores security challenges specific to quantum computing itself. We employ a mixed-methods evaluation using pre- and post-surveys to assess student learning outcomes and engagement. Results indicate significant improvements in students' understanding of quantum and hardware security, with strong positive feedback on course structure and remote instruction (mean scores: 3.33–3.83 on a 4-point scale). Additionally, students reported increased interest in careers in quantum computing and cybersecurity (M=3.67) and recognized the relevance of these skills to their future goals (M=3.5). In addition students reported several unexpected positive outcomes, such as gaining insights into industry-standard security measures and securing related internships.

\section{Introduction}
Quantum computing represents a revolutionary paradigm in computational technology, offering unprecedented capabilities to solve complex problems across various domains. Examples include machine learning \cite{cong2019quantum}, security \cite{ghosh2023primer}, drug discovery \cite{cao2018potential}, and optimization \cite{farhi2014quantum}. The integration of quantum computing and cybersecurity presents a paradigm shift that demands a comprehensive reevaluation of our approach to education and workforce preparation. In response, ensuring a secure cyberspace has been recognized as one of the National Academy of Engineering's (NAE) Grand Challenges. As quantum technologies continue to evolve at an unprecedented pace, they introduce both new vulnerabilities and opportunities in the cybersecurity landscape, necessitating awareness among researchers and developers, as highlighted by a recent workshop \cite{Workshop} hosted by the Pittsburgh Quantum Institute, supported by the National Science Foundation and the White House Office of Science and Technology Policy. This rapidly changing environment requires a workforce equipped with a nuanced understanding of both classical and quantum threat vectors, as well as proficiency in state-of-the-art defense mechanisms. Traditional cybersecurity education, often reliant on lecture-based instruction, is increasingly insufficient for preparing students to address emerging quantum threats and threats to cloud based quantum computing. Hands-on learning approaches, incorporating real-world problem-solving and industry collaboration, have been shown to enhance skill development and engagement \cite{rolston2015engineering}. In quantum security, experiential education—using quantum simulators, cloud-based hardware, and interactive security exercises—plays a crucial role in bridging the gap between theoretical knowledge and practical application.

The advent of quantum computing has sparked widespread discussion about its potential to solve classically intractable problems like integer factorization, thereby compromising traditional cryptographic systems. However, comparatively little attention has been given to the intrinsic cybersecurity challenges of quantum computing systems themselves. Unlike classical architectures, quantum systems face unique vulnerabilities rooted in the laws of quantum physics—such as decoherence (environmentally induced loss of fragile quantum states), crosstalk (unintended qubit interactions corrupting computations), and hardware supply chain risks (e.g., compromised superconducting materials or maliciously altered control electronics). These threats demand quantum-specific safeguards, such as error-correction protocols resilient to decoherence, dynamic qubit isolation to mitigate crosstalk, and hardware authentication for cryogenic components—none of which are addressed by classical cybersecurity frameworks. Current cybersecurity curricula, designed for digital logic and silicon-based systems, lack the tools to address these challenges, as they operate under fundamentally distinct assumptions. For instance, classical approaches to side-channel attacks or tamper detection fail to account for quantum superposition, entanglement, or the analog control pulses governing qubits. As quantum infrastructure advances, this gap leaves critical vulnerabilities unaddressed.

Several previous works have addressed the topic of quantum education and workforce development in recent years. Hughes et al. \cite{hughes2022assessing} conducted a survey of quantum industry needs, identifying a diverse range of job opportunities that require a combination of quantum-specific and general technical skills. Similarly, Aiello et al. \cite{aiello2021achieving} addressed challenges in developing quantum information science and engineering (QISE) education programs. Their work underscored the importance of incorporating industry input, providing hands-on training, and fostering broader participation. They advocated for inclusive curricula that emphasize essential skills such as programming, laboratory techniques, and quantum fundamentals, aligning these with industry requirements. Kaur and Venegas-Gomez \cite{kaur2022defining} offered a comprehensive review of global quantum education initiatives, including academic degree programs, online courses, workshops, and community-driven events. Their study highlighted the necessity of diverse educational pathways to develop the quantum workforce, ranging from formal academic qualifications to specialized industry training programs. However, these studies lack significant discussion of cybersecurity aspects within quantum education. While quantum technologies are poised to significantly impact cryptography and information security, current educational initiatives often neglect cybersecurity considerations. As quantum systems become increasingly prevalent, it is imperative to incorporate security topics into quantum curricula. Doing so will ensure that the future workforce is equipped to tackle emerging cyber threats and seize opportunities in the quantum era. This necessitates a dedicated focus on quantum cybersecurity within educational frameworks, equipping future experts with the knowledge and tools to navigate the complexities of securing quantum technologies. 

In response to these challenges and building upon successful educational models, we present EE-597: Introduction to Hardware Security, a pioneering curriculum explicitly designed to address the cybersecurity of quantum and classical computing systems. The course integrates practical learning through simulations, cloud-based access to quantum hardware, and industry-relevant tools, seamlessly combined with foundational concepts in classical security. 

To the best of our knowledge, this is the first comprehensive attempt to advance quantum cybersecurity education through a hands-on, activity-based approach that integrates both classical and quantum concepts. In future iterations of the course, we plan to further enhance the learning experience by implementing additional pedagogical techniques and extending the scope by offering a similar much advanced course at the graduate level. As we continue to refine and expand this curriculum, we anticipate it will play a crucial role in shaping the future of quantum cybersecurity education.

The rest of the paper is organized as follows: Section 2 outlines the proposed hardware security curriculum in detail. Section 3 describes the piloting and evaluation strategy. Section 4 explores potential avenues for future adoption of the curriculum. Finally, Section 5 presents the conclusions.

\section{Novel Hardware Security Curriculum}

We present the implementation and evaluation of a novel hardware security course that integrates hands-on quantum security learning experiences, combining simulations and cloud-based access to quantum hardware with classical security concepts.

\subsection{Curriculum Objectives}

The primary objectives of this curriculum are:  

a) To bridge theoretical concepts with practical, industry-relevant skills in classical hardware security, and quantum computing, ensuring a comprehensive understanding of secure systems.

b) To enhance student's understanding of security challenges and mitigation strategies across classical software, hardware, and quantum domains through experiential and project-based learning.

c) To establish a robust model for teaching secure computing principles, integrating classical hardware security with emerging quantum technologies and post-quantum cryptography.

d) To create a multidisciplinary approach that combines quantum computing, cryptography, and cybersecurity to provide a holistic understanding of the challenges and opportunities in secure computing.

e) To equip students with the skills to analyze vulnerabilities in classical hardware and quantum systems, and design robust solutions for secure communication and computation.

f) To foster interdisciplinary collaboration by combining principles from electrical engineering, computer science, and quantum physics in the curriculum.

g) To prepare students for cutting-edge research or industry roles by building expertise in secure system design, hardware security techniques, and quantum cybersecurity protocols.

h) To address ethical considerations in designing secure systems, ensuring the responsible development of both classical and quantum technologies.

\subsection{Course Structure}

The course structure is designed to address classical, hardware, and quantum security comprehensively while maintaining a balance between theoretical learning and hands-on application. The details are as follows:

\subsubsection{\textbf{Lectures}}
The lecture series began with foundational concepts in hardware security, covering Physically Unclonable Functions (PUFs) and True Random Number Generators (TRNGs) as key cryptographic primitives for device authentication. Students explored advanced topics such as hardware Trojans, supply chain vulnerabilities, and counterfeit detection mechanisms, with an emphasis on their implications for secure computing environments. Field Programmable Gate Arrays (FPGA) and Application-specific integrated circuit (ASIC) security were also covered in depth, introducing students to threats posed by side-channel attacks, including power analysis and electromagnetic (EM) leakage, along with countermeasures such as trusted execution environments (TEEs), and advanced metering techniques for anti-piracy protection.

In the quantum computing segment, students were introduced to the fundamentals of quantum information encoding and circuit design before progressing to security-specific challenges. Key topics included the vulnerabilities of quantum devices due to decoherence, crosstalk-induced errors in multi-tenant quantum processors, and potential risks posed by malicious quantum transpilers that modify circuit structures, highlighting potential backdoors in quantum transpilation that could manipulate computations at a low level, compromising results while remaining undetectable at the algorithmic level.

\subsubsection{\textbf{Hands-on Activities}}

The hands-on activities geared towards quantum security provided students with practical experience in quantum computing and its security implications. These activities were designed to give participants a deep understanding of quantum circuit design, execution on real quantum hardware, and exploration of quantum security vulnerabilities.

\textbf{a) IBM Quantum Resources and Circuit Execution:} Students were introduced to IBM's (one of the leading developers of quantum computers and software toolchain) quantum computing resources through guided lectures and tutorials. These sessions covered:

- IBM Quantum Composer: A graphical tool for building quantum circuits by dragging and dropping operations \cite{lehka2022hardware}. Students learned how to: 1) Visualize qubit states using interactive q-spheres and histograms; 2) Generate OpenQASM or Python code automatically from their circuits; 3) Customize their workspace for optimal circuit design.

- Circuit Execution on Real Quantum Hardware: Participants ran their designed circuits on actual IBM quantum systems, gaining insights into: 1) The effects of device noise on quantum computations; 2) Differences between simulated results and those from real quantum hardware; 3) The process of submitting jobs to quantum backends.

\textbf{b) Quantum Circuit Design and Implementation:} Building on their understanding of IBM's tools, students engaged in hands-on exercises to design and implement quantum circuits:

- Circuit Construction: Using IBM Quantum Composer, participants learned to: 1) Apply single-qubit gates (e.g., Hadamard, NOT) and multi-qubit gates (e.g., CNOT); 2) Measure qubit states and interpret results; 3) Optimize circuit depth and gate count for better performance.

- OpenQASM Programming: Students were introduced to OpenQASM 2.0, learning to: 1) Write quantum circuits using code instead of graphical interfaces; 2) Define custom operations for more complex quantum algorithms; 3) Understand the relationship between graphical circuit design and code-based approaches.

\textbf{c) Exploring Quantum Security Vulnerabilities:} A significant portion of the hands-on activities focused on investigating quantum security vulnerabilities, with a particular emphasis on various attack models and defense.

- Crosstalk Attacks and Practical demonstration: Students explored the concept of crosstalk in quantum systems, understanding its potential for malicious exploitation \cite{ash2020analysis}. This module focused on: 1) How crosstalk can be leveraged for attacks in multi-tenant quantum environments where multiple user programs share the same hardware; 2) The impact of crosstalk on the reliability of quantum computations; 3) Participants engaged in exercises simulating crosstalk attacks using a python script and cloud based access to the quantum hardware.

\textbf{d) Quantum PUF (QuPUF) Design and implementation:} Students also got the opportunity to implement quantum PUF \cite{phalak2021quantum} for fingerprinting various quantum hardware. This session focused on:

- Hello World: The Hello World program of creating a two-qubit bell state circuit was implemented by the participants using 1) IBM Quantum Composer and 2) Python code with the help of Qiskit library, helping them overall to grasp the basics of programming in quantum computing.

- Superposition-based QuPUF implementation: Participants created the superposition-based QuPUF and ran it in both simulation and on real quantum hardware. This helped the students understand the differences between simulation and actual hardware runs, thereby helping them understand the underlying working of superposition-based QuPUF.

- Decoherence-based QuPUF implementation: Similar to superposition QuPUF, students implemented the decoherence-based QuPUF and ran it in both simulation and hardware. The differences in results of both cases helped the students understand the working of decoherence-based QuPUF.

\subsubsection {Case Studies}
Examination of real-world hardware security breaches, and quantum security scenarios, with discussions on lessons learned and potential solutions \cite{phalak2021quantum}\cite{castro2006securing}\cite{azab2014hypervision}\cite{sgxsgxpectre}\cite{ash2020analysis}\cite{burow2017control}.

\subsubsection{Guest Lectures and Industry Collaboration}

The course featured valuable insights from both academic and industry experts, fostering a comprehensive understanding of cutting-edge topics in quantum computing and security. A faculty member from the Nanyang Technological University (NTU) delivered an engaging lecture on quantum threat mitigation strategies, emphasizing the critical role of secure quantum algorithms in the evolving cybersecurity landscape. Complementing this, an industry expert from Intel shared practical perspectives on emerging challenges, and trends in the VLSI industry, providing students with a real-world context for the theoretical concepts discussed in class.

\subsubsection{Weekly Quizzes}
Short, targeted assessments to reinforce foundational concepts in classical and quantum security, as well as hardware security techniques.

\subsubsection{Capstone Projects}
Interdisciplinary projects requiring students to identify and address vulnerabilities in classical and quantum systems, with deliverables including secure hardware designs or quantum-enhanced cryptographic protocols.

\subsubsection{Flipped Classroom Approach}
Encouraging active participation and deeper understanding by assigning preparatory materials for pre-class study, followed by in-class collaborative problem-solving.

\begin{figure*}
    \centering
    \includegraphics[width= 5.5in]{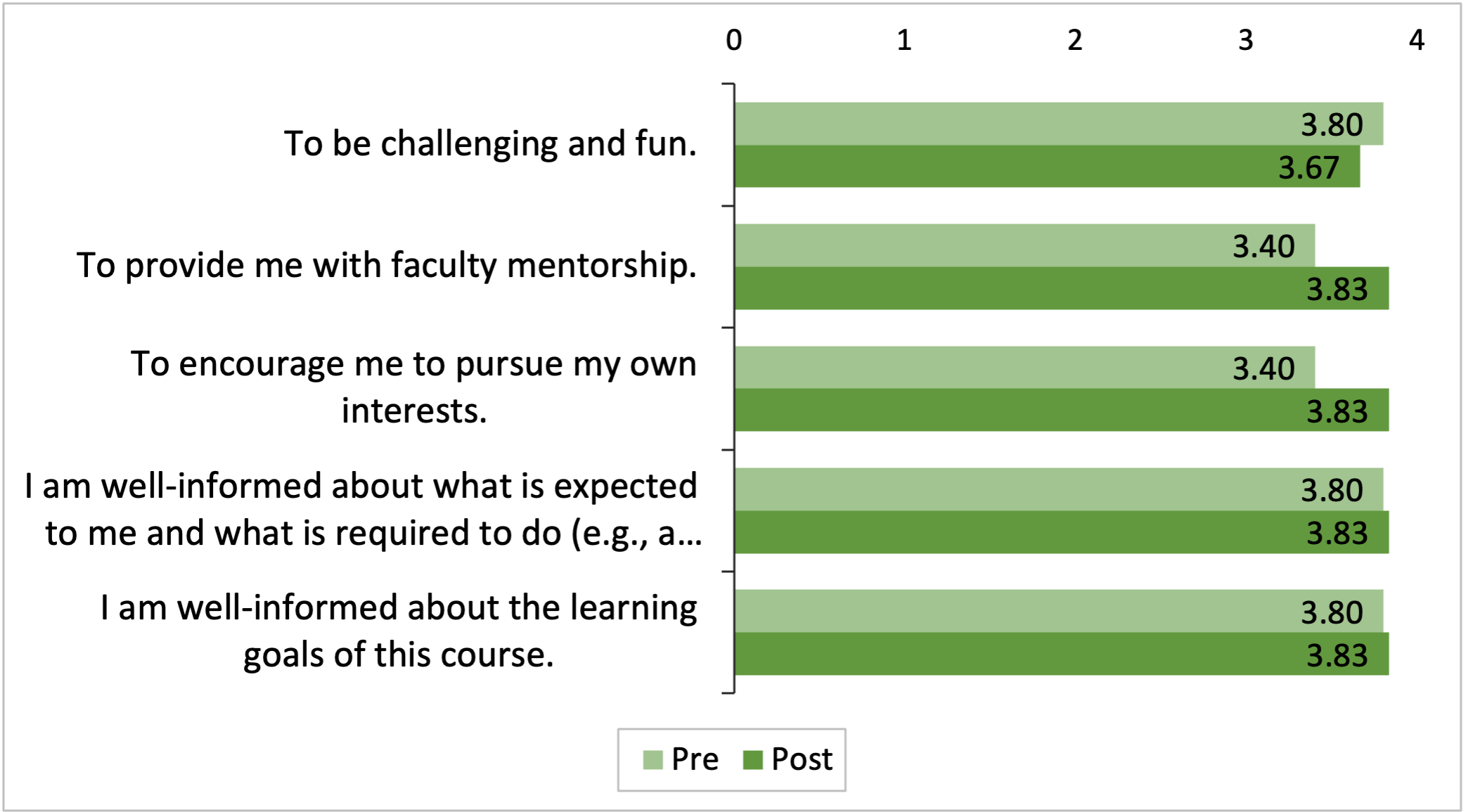}
    \caption{Comparison of student’s expectations before and after the course. The expectations were measured on a scale from 1 to 4 (1 = strongly disagree to 4 =
strongly agree).}
    \label{fig1}
\end{figure*}

\section{Implementation and Piloting}

We evaluated the impact of a quantum computing curriculum on graduate student learning, specially focusing on the developed course - EE597: Introduction to Hardware Security. This course features a hands-on curriculum on quantum security, combining simulations with cloud-based access to actual quantum hardware, aimed at enhancing the understanding and awareness of quantum computing issues among students in CS and STEM fields.

The key research questions addressed in this evaluation included:

1. Were the planned activities executed successfully to ensure learning progress?

2. How has the student's knowledge of quantum computing and cybersecurity concepts and skills improved?

3. How has the student's understanding of quantum computing and cybersecurity issues evolved?

4. How has the interest of students in quantum computing, cybersecurity, and related careers changed?

The course evaluation was conducted using pre- and post-surveys. The student participants were asked to participate in the pre-survey at the beginning of the semester and the post-survey right after the semester was over. However, a focus group could not be conducted as planned due to insufficient voluntary participation. Dr. Kathleen Hill and Dr. Jungeun Lee from the Penn State Center for Science and the Schools (CSATS) conducted the data analysis.

\subsection{Demography}

During the Spring 2024 semester, seven graduate students participated in the course. Five of the seven students completed the pre-survey, which included basic demographic questions. Therefore, the demographic data presented here pertains only to those who completed the pre-survey. Among the five respondents, 60\% were female and 40\% were male. All participants were Asian graduate students. Four of the five students were studying Electrical Engineering, while one was studying Computer and Technology.

\begin{figure*}
    \centering
    \includegraphics {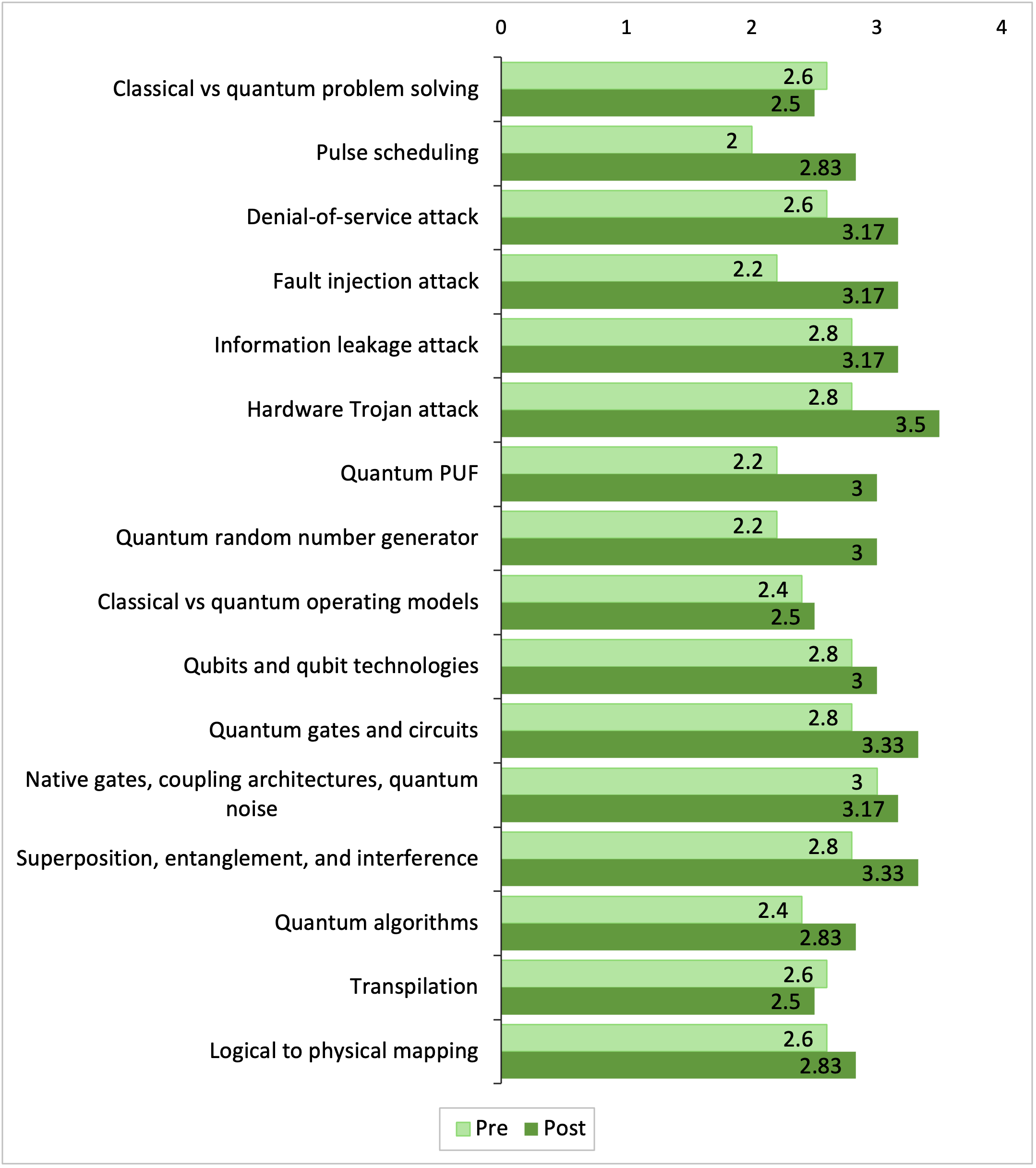}
    \caption{Student's familiarity of knowledge on topics from quantum computing and security before and after the course.}
    \label{Fig2}
\end{figure*}

\subsection{Expectation}

To assess whether the planned activities were executed successfully to ensure learning progress, we asked students about their expectations for this course and whether these expectations were met at the end of the course. Fig.\ref{fig1} displays the comparison of student's expectations before and after the course. The expectations were measured on a scale from 1 to 4 (1 = strongly disagree to 4 = strongly agree). Although there was a slight decrease in perceived challenge and fun, significant increases were noted in faculty mentorship and encouragement to pursue personal interests, indicating that the course exceeded expectations in these areas. There were also slight improvements in understanding course requirements and clarity of learning goals. Overall, the course met or exceeded student's expectations, particularly in mentorship and encouraging personal interests.

\subsection{Assessment of Student Knowledge Improvements}

To assess how student's knowledge of quantum computing and cybersecurity concepts and skills improved, and how their understanding of these issues evolved, we asked participants about their goals and achieved outcomes. Additionally, students were asked to indicate their familiarity with the topics discussed.

\subsubsection{Goals and outcomes}

At the start of the course, students were asked to share their main goals and expectations. A common theme in their responses was the desire to enhance their knowledge of hardware and quantum security, with a focus on addressing security vulnerabilities and learning secure design methods. Many students also expressed interest in understanding how to defend against physical and logical attacks while keeping up with advancements in emerging technologies. Following is the list of student responses:

a) To learn about the hardware security and how to mitigate the problems regarding the security.

b) Learn about hardware security and quantum security.

c) I work in the domain of quantum hardware security; I hope to learn more about this subject so that I feel confident while doing research in this field hereafter.

d) The thing I would most like to achieve in this hardware security course is gaining a broad understanding of the vulnerabilities in modern digital systems and the latest techniques for building more secure hardware. I want to learn how to integrate security starting from the hardware design phase, rather than treating it as an afterthought. By the end of the course, my goal is to have an awareness of physical attacks like side-channel analysis and invasive probes, as well as logical attacks like hardware Trojans, so that I can design defenses against them. I also hope to stay up-to-date on emerging technologies like non-volatile memories and quantum computing and understand their security implications. Most importantly, I aim to adopt a security mindset that allows me to identify risks and create trusted, tamper-resistant hardware and embedded systems.

e) Understanding the current state of hardware security, and existing vulnerabilities.

After the course was concluded, the student participants were asked about their learning goals and achieved outcomes from the course. The students reported achieving their goals in understanding hardware security concepts, learning about quantum computing, and applying knowledge to their research. Weekly quizzes helped focus their studies. They gained insights into attacks like trojans and side channels, and the connection between hardware and software security. Some wrote research papers, and others appreciated learning industry-level practices and implementation techniques. Overall, the course successfully enhanced their knowledge in hardware security and quantum computing. Following is the list of student responses: 

a) I aimed to get familiar with the concepts related to security in hardware domain which this course helped me to understand. Got introduced to quantum computing. weekly quiz helped to focus on the topics covered.

b) I expected to learn about hardware security and use it in my research towards my PhD.

c) I ended up writing a paper using that knowledge that I hope to submit to a conference soon.

d) 1. Quantum compute encryption 2. JTAG is used to do chip sanity check 3. Multiple encryption algorithms implementations.

e) I learned more about hardware and the thought process for securing it - types of attacks an adversary can do like trojans, side channels - understand the connection between hardware and software.

f) 1. Exposure to importance of hardware security and interesting implementation techniques 2. Learnt some interesting cyber issues and securities that can be done on hardware at a chip design level 3. Understanding of industry level of the domain.

g) 1) I aimed to and learned about fundamentals of security and hardware security. 2) I aimed to learn about basics of quantum computing. I got a basic hands-on on quantum computing. 3) I aimed to learn about various security vectors regarding VLSI and quantum computing, got an exposure for those in this course.

\begin{figure*}
    \centering
    \includegraphics{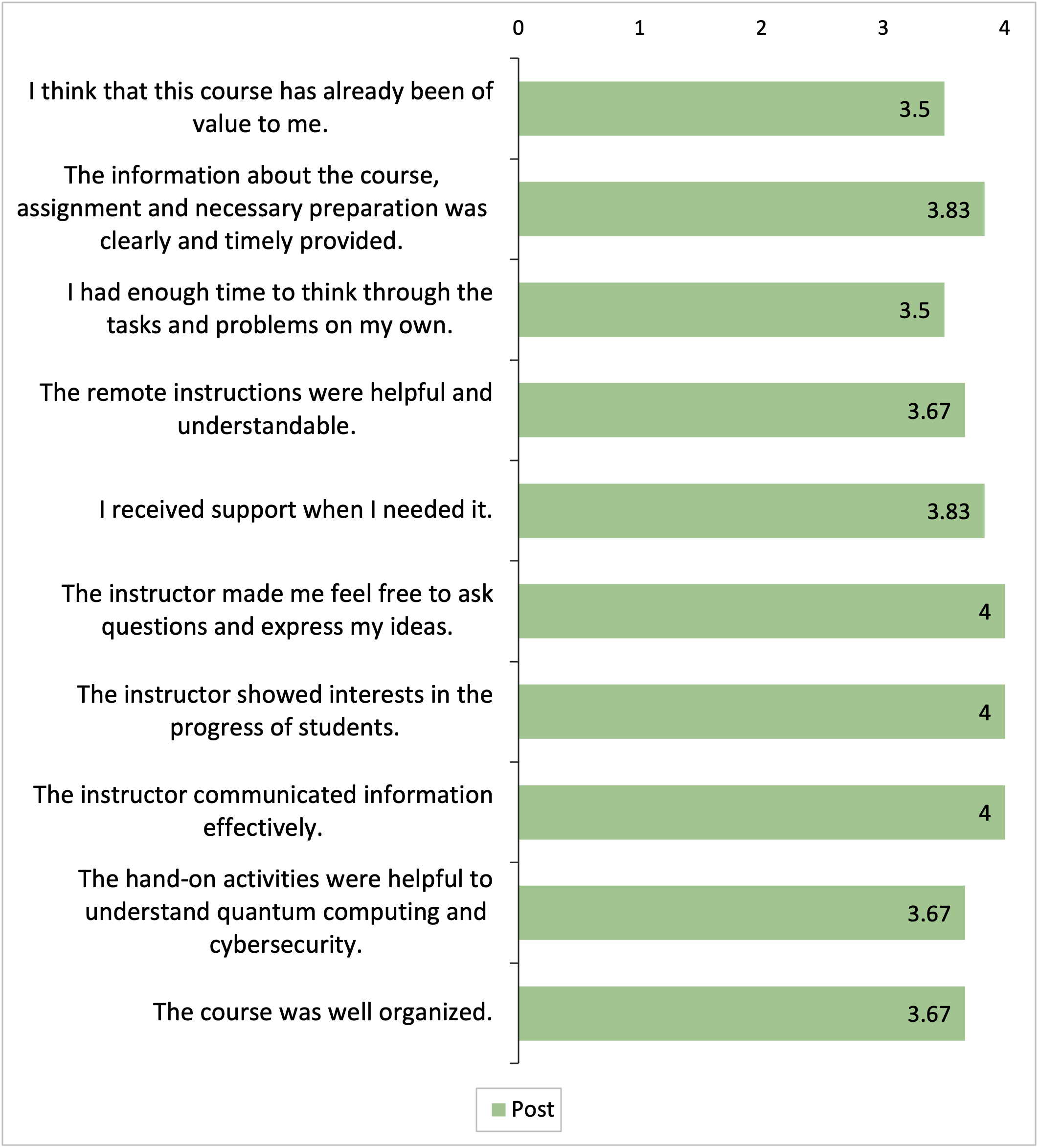}
    \caption{Post-course student feedback ratings (1–4 scale) on course quality.}
    \label{Fig3}
\end{figure*}

\begin{figure*}
    \centering
    \includegraphics[width= 4.5 in]{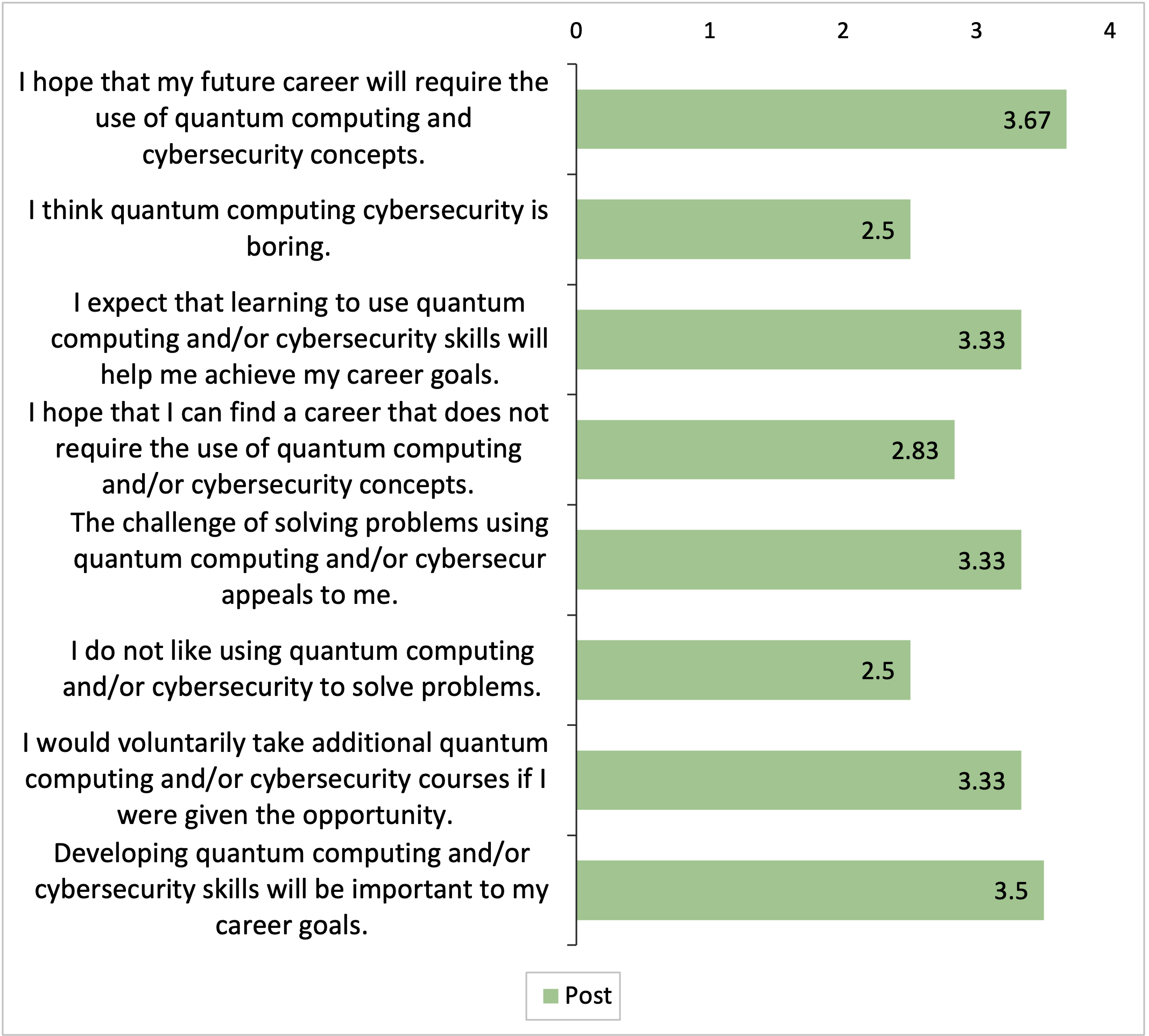}
    \caption{Student's attitudes toward the discipline.}
    \label{Fig4}
\end{figure*}

Most students successfully achieved their goals in the course, with one student noting that the inclusion of physical chip design and security techniques would have been beneficial. Beyond their initial objectives, students reported several unexpected learning outcomes, including an improved understanding of quantum computing and software security challenges, exploration of industry-standard hardware security measures, and the generation of new ideas for further research. Notably, one student secured an interview for a related internship, while another found the course exceeded their expectations, providing valuable insights beyond their original goals. Overall, the course effectively enhanced students’ knowledge of hardware security and quantum computing. They also appreciated the exposure to unexpected learning opportunities and real-world industry practices.

\subsubsection{Familiarity with knowledge}

Student participants were asked to indicate their familiarity on topics related to the course at the beginning and at the end of the course, which was measured on a scale from 1 to 4 (1 = not at all familiar to 4 = highly familiar). Fig.\ref{Fig2} displays the comparison of student's familiarities on knowledge before and after the course, which illustrates the improvement in student's familiarity with various topics in quantum computing and cybersecurity after completing the course. Key areas of significant progress include pulse scheduling, denial-of-service attacks, fault injection attacks, hardware Trojan attacks, and quantum PUF, with post-course familiarity ratings notably higher than pre-course ratings. Students also showed increased understanding of various topics including quantum gates and circuits, superposition, entanglement, and quantum algorithms. Overall, the course enhanced student's knowledge and skills across a broad range of topics.

\subsection{Overall Experience}

At the end of the course, student participants were asked to rate their overall experience in the course on a scale from 1 to 4 (1 = strongly disagree to 4 = strongly agree). Fig. \ref{Fig3} shows the feedback on various aspects of the course. Overall, students found the course valuable (M=3.5) and well-organized (M=3.67). They appreciated the clarity and timeliness of information provided (M=3.83). Remote instructions were helpful (M=3.67), and students received support when needed (M=3.83). They also appreciate the instructor’s support. Hands-on activities seemed to be helpful (M=3.67) in understanding quantum computing and cybersecurity.

\subsection{Attitudes towards the discipline}

The student responses reflect a strong positive attitude towards the discipline, as measured on a scale from 1 to 4 (1 = strongly disagree to 4 = strongly agree) (Fig.\ref{Fig4}). Most students expressed a desire for future careers involving quantum computing and cybersecurity (M=3.67) and believed these skills would help them achieve their career goals (M=3.33). They recognized the importance of these skills for their professional aspirations (M=3.5) and showed a willingness to pursue additional courses in these areas (M=3.33). However, there was moderate agreement that some students might pursue careers not requiring these skills (M=2.83), and a few found the subjects less interesting or particularly challenging (M=2.5). Overall, the responses indicate a strong interest and commitment to integrating quantum computing and cybersecurity into their future professional endeavors.

\section{Future Adoption}

Building on the success of the current implementation, the curriculum for EE-597 will be adapted to expand enrollment, enhance accessibility, and maintain relevance in the evolving landscape of quantum and cybersecurity. The small sample size (N=7) in the current study limits the generalizability of findings, so future offerings will aim for a larger cohort. This will be achieved by (i) circulating promotional flyers within EE, CS and physics departments, (ii) reaching out to quantum and security clubs at Penn State and senior undergraduate students and (iii) making announcements about this course in classes taught on quantum computing and software security. The curriculum will be enhanced to include more industry talks and career counseling in the quantum and security areas. The initial class size is expected to be 30-40 students, ensuring interactive learning while maintaining accessibility and allowing for more comprehensive assessment of student learning outcomes and engagement across diverse backgrounds. To improve inclusivity, targeted outreach efforts will encourage participation from underrepresented groups, including women, minorities, and first-generation students. Collaborations with diversity-focused organizations, such as Women in Engineering and cybersecurity clubs, will help promote awareness. Additionally, preparatory modules covering essential background knowledge will be offered to support students with varying levels of prior experience. The course will continue to emphasize hands-on learning and collaborative research projects. Recorded lectures will be provided before each session, allowing in-class time to focus on discussions and practical applications. This flipped-classroom approach will maximize engagement with complex quantum security topics. Group-based activities and projects using quantum simulators and cloud-based quantum hardware will reinforce learning while fostering teamwork and communication skills. Regular feedback through surveys and focus groups will guide iterative improvements ensuring they remain impactful and inclusive.


\section{Conclusion}

This work highlights the urgent need for an educational shift to address the dual challenges of classical and quantum cybersecurity. By implementing a novel curriculum that integrates hands-on quantum and hardware security training with classical concepts, we demonstrated a practical and impactful approach to bridging theoretical knowledge and industry-relevant skills. The positive outcomes, including enhanced student understanding, increased interest in cybersecurity careers, and practical exposure to real-world scenarios, underscore the curriculum's effectiveness. As quantum computing continues to evolve, such educational initiatives will be critical in preparing a skilled workforce capable of navigating the complex security landscape of the future.

\vspace{4\baselineskip}\vspace{-\parskip} 
\footnotesize 
\bibliographystyle{unsrtnat} 
\bibliography{ASEEpaper}


\end{document}